\documentclass[twocolumn,showpacs,showkeys,preprintnumbers,amsmath,amssymb]{revtex4}
\usepackage{graphicx}% Include figure files
\usepackage{dcolumn}% Align table columns on decimal point
\usepackage{bm}% bold math

\begin{document}

%\preprint{APS/123-QED}

\title{Long-term Correlations and $1/f^\alpha$ Noise in the Steady States 
\\of Multi-Species Resistor Networks}

\author{C. Pennetta}
\affiliation{Dipartimento di Ingegneria dell'Innovazione, 
Universit\`a del Salento and Consorzio Nazionale Interun. 
per le Scienze Fisiche della Materia (CNISM), Via Arnesano, I-73100, 
Lecce, Italy.} 
\email{cecilia.pennetta@unile.it}

\author{E. Alfinito}
\affiliation{Dipartimento di Ingegneria dell'Innovazione, 
Universit\`a del Salento and Consorzio Nazionale Interun. 
per le Scienze Fisiche della Materia (CNISM), Via Arnesano, I-73100, 
Lecce, Italy.}

\author{L. Reggiani}
\affiliation{Dipartimento di Ingegneria dell'Innovazione, 
Universit\`a del Salento and Consorzio Nazionale Interun.
per le Scienze Fisiche della Materia (CNISM), Via Arnesano, I-73100, 
Lecce, Italy.}

\date{\today}

\begin{abstract}
We introduce a multi-species network model which describes the resistance
fluctuations of a resistor in a non-equilibrium stationary state. More 
precisely, a thin resistor characterized by a $1/f^\alpha$ resistance noise 
is described as a two-dimensional network made by different species of 
elementary resistors. The resistor species are distinguished by their 
resistances and by their energies associated with thermally activated 
processes of breaking and recovery. Depending on the external conditions, 
stationary states of the network can arise as a result of the competition 
between these processes. The properties of the network are studied 
as a function of the temperature by Monte Carlo simulations carried out in 
the temperature range $300 \div 800$ K. At low temperatures, the resistance 
fluctuations display long-term correlations expressed by a power-law behavior 
of the auto-correlation function and by a value $\approx 1$ of the 
$\alpha$-exponent of the spectral density. On the contrary, at high 
temperatures the resistance fluctuations exhibit a finite and progressively 
smaller correlation time associated with a non-exponential decay of 
correlations and with a  value of the $\alpha$-exponent smaller than one. 
This temperature dependence of the $\alpha$ coefficient reproduces 
qualitatively well the experimental findings. 
\end{abstract}

\maketitle

%\begin{multicols} {2}

\section{Introduction}
Since a long time the analysis of the resistance fluctuations has turned out 
to be a very powerful tool for probing various condensed matter systems 
\cite{review,upon05} and, in particular, disordered materials 
\cite{review,upon05,torquato,sahimi,stauffer} like conductor-insulator 
composites \cite{bardhan}, granular systems \cite{grier}, porous \cite{bloom} 
or amorphous materials \cite{kakalios,alers,weissman_1f} and organic 
conducting blends \cite{planes_1f}. Therefore, many experimental and 
theoretical investigations have been devoted to study the properties of the
resistance noise as a function of temperature, bias strength and of the main 
material parameters 
\cite{review,upon05,bardhan,grier,bloom,kakalios,alers,weissman_1f,planes_1f,chen_1f,raychaud,chiteme,em_fn04,ciliberto}. 
Only few studies have considered the behavior of disordered materials over the
full range of bias values, by analyzing under a unified approach the linear 
regime and the nonlinear regime up to the threshold for electrical breakdown 
\cite{bardhan,sornette97,pre_fnl,pen_ng}, even though this kind of studies 
can provide significant insights about basic properties of the response of 
nonequilibrium systems  
\cite{bardhan,ciliberto,sornette97,pre_fnl,pen_ng,stan_zap,odor,derrida}.

In previous works we have introduced \cite{upon99} and developed
the Stationary and Biased Resistor Network (SBRN) model 
\cite{pre_fnl,pen_ng,pen_ng_fn04,pen_prb} which is based on a resistor 
network approach \cite{torquato,sahimi,stauffer} and which allows the study 
of the electrical conduction and noise of disordered materials over the full 
range of bias values. This model describes a disordered conducting material 
as a resistor network whose structure is determined by a stochastic 
competition between a biased percolation \cite{gingl,prl_fail} of ``broken'' 
resistors (high resistivity resistors) and a biased recovery process (which 
restores the ``regular'' resistors with ``normal'' resistivity) 
\cite{pre_fnl,upon99}. The SBRN model provides a good description of many 
features associated with the resistance fluctuations in nonequilibrium 
stationary states of disordered resistors 
\cite{bardhan,planes_1f,pre_fnl,pen_ng,pen_ng_fn04,leturcq}, with resistor 
trimming processes \cite{grimaldi} and with agglomeration and thermal 
instability phenomena in metallic films \cite{sieradzki,kim}. Further 
successful applications of the model concern the study of electrical breakdown
phenomena in composite and granular materials 
\cite{bardhan,pre_fnl,pen_ng,pen_ng_fn04}, including electromigration damage 
of metallic lines \cite{em_fn04,pen_prb,islam}.
  
However, this model only applies to systems characterized by a Lorentzian 
noise \cite{review}, i.e. by a power spectral density of the resistance 
fluctuations scaling as $1/f^2$ at high frequencies and becoming flat at 
frequencies below the so called corner value $f_c$. The existence of a corner 
frequency in the spectrum corresponds in the time domain to the existence of a
well defined characteristic time, $\tau$ (correlation time), associated with 
the decay of the correlations in the resistance fluctuations \cite{review}. In 
the particular case of Lorentzian noise this decay is exponential, i.e. 
the auto-correlation function of the resistance fluctuations is given 
by \cite{review}: 
\begin{equation}
C_{\delta R}(t) \equiv \langle \delta R(t)\delta R(t+\tau)\rangle = \langle (\Delta R)^2\rangle \exp(-t/\tau)  \label{eq:cor_exp}
\end{equation}
where $\langle(\Delta R)^2\rangle$ is the variance of the resistance 
fluctuations. On the other hand, it is well known \cite{review} 
that at low frequencies many condensed matter systems display $1/f$ resistance
noise, i.e. a spectral density scaling over several frequency decades as 
$1/f^\alpha$, with $\alpha \approx 1$. The amazing presence of $1/f$ noise in 
a large variety of natural phenomena represents a puzzling problem not yet 
fully solved. Therefore, a huge quantity of models providing signals with 
$1/f$ noise have been proposed 
\cite{review,upon05,grier,kakalios,zhang,btw,jung,kaulakys_1f,kiss,kolek,celasco_1f,rakhmanov_1f,shklovski_1f,shtengel_1f}.
In particular, the ubiquity of the $1/f$ noise has given rise to many attempts
to explain it in terms of a universal law. One simple way to obtain a $1/f$ 
spectrum is by superimposing a large number of Lorentzian spectra with an 
appropriate distribution of the correlation times: $D(\tau) \propto 1/\tau$, 
distributed over many orders of magnitude of $\tau$ values \cite{review}. 
In some cases, the distribution of the correlation times of the elementary 
processes contributing to the $1/f$ spectrum can be derived from the 
distribution of some variable on which the correlation times themselves 
depend, as in the case of the pioneering works of Mc Whorter \cite{review} and
Dutta and Horn \cite{review}. In particular, these authors proposed 
\cite{review} that the origin of the $1/f$ noise can be attributed to a 
thermally activated expression of the correlation times, associated with a 
broad distribution of the corresponding activation energies, an assumption 
physically plausible for many systems. On the other hand, many other important
contributions to the understanding of the $1/f$ noise have been advanced in 
the last twenty years, which have shown that the presence of a $1/f$ spectrum 
can also arise from other basic reasons 
\cite{upon05,grier,kakalios,zhang,btw,jung,kaulakys_1f,kiss,kolek,celasco_1f,rakhmanov_1f,shklovski_1f,shtengel_1f}. 
The dynamical random network model \cite{zhang} and other spin-glasses models 
\cite{review,zhang} provide a good explanation of the $1/f$ noise in 
conducting random magnetic materials. Dissipative self-organised criticality 
(SOC) models, started from the famous work of Bak, Tang and Wiesenfeld 
\cite{btw}, clarify the origin of $1/f$ spectra in certain dissipative 
dynamical systems naturally evolving into a critical state. Avalanche models 
\cite{jung}, clustering models \cite{kaulakys_1f} and percolative models 
\cite{grier,kakalios,kiss,kolek,celasco_1f} represent other relevant classes 
of theoretical approaches explaining the appearence of $1/f$ noise in a 
variety of systems. Thus, the conclusion, now largely accepted in the 
literature \cite{review,upon05,celasco_1f}, is that a unique, universal 
origin of $1/f$ does not exist, even though classes of systems can share a 
common basic origin of $1/f$ noise.  

The aim of this paper is to present a new model which includes and extends
to systems characterized by $1/f^{\alpha}$ noise the main feature of the SBRN 
model, i.e. the stochastic competition between thermally activated and biased 
processes of breaking and recovery of the elementary resistors of a network
\cite{pre_fnl}. Actually, this feature has turned out to be essential 
for the success of the SBRN model in the description of the wide phenomelogy 
associated with the resistance fluctuations of disordered resistors, in both 
nonequilibrium stationary states and nonstationary failure states 
\cite{bardhan,planes_1f,pen_prb,leturcq,grimaldi,sieradzki,kim}.
To this purpose, here we study a network made by several species of resistors, 
where each species is characterized by a resistance value and by a pair of 
activation energies associated with the breaking and the recovery processes. 
For this reason we call this model ``multi-species network'' (MSN) model. 
The states, either stationary or nonstationary, of the MSN result from the 
stochastic competition between these processes of breaking and recovery 
involving the elementary resistors. In analogy with the Dutta and Horn model 
\cite{review}, we take the breaking and recovery activation energies 
distributed in a broad range of values. A choice physically justified in 
disordered systems, like granular and amorphous materials, composites, etc. 
where the orientational disorder present inside these materials can give rise 
to different energy barriers for the electron flow along the different 
conducting paths \cite{review,torquato,sahimi,zhang}. As a result, the 
resistance fluctuations of the MSN exhibit a $1/f^{\alpha}$ spectrum, where 
the value of $\alpha$ depends on the external conditions (temperature and 
electrical bias). In particular, in this paper, we discuss the states of 
a MSN as a function of the temperature and in the linear regime of the 
external bias, when Joule heating effects are negligible. 

The paper is organised as follows: in Sect. II we illustrate the MSN model, 
while in Sect. III we report the results by focusing in particular on the 
role played by the temperature on the fluctuations properties of the 
multi-species network. Finally, in Sect. IV we draw the conclusions of 
this study. 

\section{Model}
According to the SBRN model \cite{pre_fnl,pen_ng} a conducting thin film with 
granular structure is described as a two-dimensional square-lattice resistor 
network \cite{torquato,stauffer}. For simplicity here we study a network 
$N \times N$ (where $N$ is the linear size of the network), even if 
networks with a different geometry can be considered  \cite{pen_prb}. The 
network is biased by an external constant current, $I$, applied through 
perfectly conducting bars placed at the left and right hand sides and it lies 
on an insulating substrate at a given temperature $T$, which acts as a thermal
bath. Each resistor can be in two different states: 
i) regular, corresponding to a resistance $r_n =r_0[1+\alpha_T(T_n - T)]$, 
where $\alpha_T$ is the temperature coefficient of the resistance and $T_n$ 
is the local temperature and ii) broken, corresponding to an effectively 
``infinite'' resistance, $r_{OP} = 10^9 r_n$. Resistors in the broken state 
will be called defects. The temperature $T_n$ is expressed as 
\cite{prl_fail}:
\begin{equation}
T_{n}=T + A[ r_{n}i_{n}^{2}+{D\over N_{neig}}\sum_l( r_{l} i_{l}^2-r_n i_n^2)]
\label{eq:bias}
\end{equation}
where $D=3/4$, the sum is performed over the $N_{neig}$ nearest neighbors of 
the n-{\em th} resistor and $i_{n}$ is the current flowing in it. The above 
expression of $T_n$ takes into account the Joule heating of the 
n-{\em th} resistor \cite{gingl,prl_fail,sornette92} and the thermal exchanges 
with its neighbors \cite{prl_fail}. The importance of both these 
effects is controlled by the parameter $A$, thermal resistance, which 
describes the heat coupling of each resistor with the substrate 
\cite{pre_fnl,gingl,sornette92}. We note that by adopting 
Eq.(\ref{eq:bias}) we are assuming for simplicity an instantaneous 
thermalization of each resistor and we are neglecting time dependent effects 
discussed in Ref. [\onlinecite{sornette92}]. 

In the initial state of the network all the resistors are taken to be 
identical: $r_n \equiv r_0$. The SBRM model then assumes that the evolution 
of the network is determined by the competition between two biased stochastic 
processes: one of breaking and the other of recovery. The former process 
consists in the transition of an elementary resistor from the regular to 
the broken state and it occurs with a thermally activated probability 
\cite{gingl}: 
\begin{equation}
W_{Dn}=\exp(-E_D/k_B T_n) 
\label{eq:wd}
\end{equation}
the latter process consists in the reverse transition and it occurs with 
probability  \cite{pre_fnl,upon99}:
\begin{equation}
W_{Rn}=\exp(- E_R/k_B T_n)
\label{eq:wr}
\end{equation}
where $E_D$ and $E_R$ are the activation energies of the two processes and 
$k_B$ the Boltzmann constant. The time evolution of the network is then 
obtained by Monte Carlo simulations which update the network resistance after 
a sweep of breaking and recovery processes, according to an iterative 
procedure detailed in Ref. [\onlinecite{pre_fnl}]. The sequence of successive 
network configurations provides a resistance signal, $R(t)$, after an 
appropriate calibration of the time scale. Then, depending on the stress 
conditions ($I$ and $T$) and on the network parameters (size, activation 
energies, $r_0$, $\alpha_T$ and $A$), the network either reaches a steady 
state or undergoes an irreversible electrical failure 
\cite{pre_fnl,pen_ng,pen_ng_fn04,pen_prb}. 
This latter possibility is associated with the achievement of the percolation 
threshold, $p_c$, for the fraction of broken resistors \cite{stauffer}. 
Therefore, for a given network at a given temperature, a threshold current 
value, $I_B$, exists above which electrical breakdown occurs \cite{pre_fnl}. 
For values of the applied current below this threshold, the steady state of 
the network is characterized by fluctuations of the fraction of broken 
resistors, $\delta p$, and of the resistance, $\delta R$, around their 
respective average values $\langle p \rangle$ and $\langle R \rangle$. 

The network described above, as provided by the SBRN model, is made by a 
single species of resistors, all characterized by the same values of 
$r_0$, $\alpha_T$, $E_D$ and $E_R$. Thus, we can speak of a ``single-species 
network''. The disorder inside this network only arises from the fact that 
some resistors can be in the broken state. As a consequence, current and 
temperature are not homogeneously distributed \cite{pen_prb,gingl}, thus 
implying inhomogeneous probabilities of breaking and recovery and, through 
the temperature coefficient $\alpha_T$, different resistance values of the 
regular resistors. Furthermore, by analogy with generation-recombination noise
in a two-levels system, this model describes a system characterized by a 
single time scale and thus by Lorentzian noise. To overcome the above 
limitations here we consider a network made by $N_{spec}$ species of 
resistors, i.e. a ``multi-species network'' (MSN). Each species is 
characterized by different values of the parameters: $r_{0,i}$, 
$\alpha_{T,i}$, $E_{D,i}$ and $E_{R,i}$ with $i=1,...N_{spec}$. Then, 
an ``identity'' (an index $i$) is attributed at random to each resistor. 
In the initial state of the network the species are taken to be present with 
the same number of resistors. As a reasonable choice, we have taken $r_{0,i}$ 
uniformly distributed inside a given range of resistance values: 
$r_{0,i} \in [r_{min},r_{max}]$, while other options are as well reasonable. 
We limit the present study to the linear regime in the external bias, when 
Joule heating effects are negligible, i.e. $T_n \equiv T$, $\forall n$ and 
consequently $r_{n,i} \equiv r_{0,i}$, $\forall n$ and $\forall i$. 
Alternatively, we can say that we study a system characterized by negligibly 
small values of the thermal resistance and of the resistance temperature 
coefficients: $A=0$ and $\alpha_{T,i}=0$, $\forall i$. We stress that the 
linear regime condition corresponds to uniform probabilities of breaking and 
recovery for the resistors belonging to each given species 
\cite{prl_fail,prl_stat}. Therefore, the results that we will report in 
Sect. III directly generalize the results of Ref. [\onlinecite{prl_stat}] to 
systems characterized by $1/f^\alpha$ resistance noise. The limitation to the 
linear regime of currents has been adopted here for simplicity and also 
because we aim to focus on the temperature dependence of the correlation 
properties of the resistance fluctuations. However, we underline that there 
is no special difficulty in extending this study to the nonlinear regime in
the applied bias. In this respect, it must be noted that we expect that all 
the other significant features associated with the SBRN model (existence of 
a threshold current for breakdown \cite{pre_fnl,pen_ng}, dependence of 
$\langle R \rangle$ and $\langle (\Delta R)^2 \rangle$ on the bias 
\cite{pre_fnl}, times to failure distribution \cite{pen_prb}, non-Gaussianity 
of the resistance fluctuations \cite{pen_ng,pen_ng_fn04}, role of the network 
geometry \cite{pen_prb}, etc.) remain unchanged for a MSN network.
 
The values of the activation energies $E_{D,i}$ and $E_{R,i}$ for the different
resistor species have been chosen as follows. In Ref. [\onlinecite{prl_stat}] 
it was proved that for a single-species network, when the average fraction of 
defects $\langle p \rangle$ is sufficiently far from the percolation threshold
($p_c=0.5$ for a $N \times N$ square-lattice network \cite{stauffer}), the 
correlation time of the resistance fluctuations is approximated by: 
%
%\vspace*{-0.2cm}
\begin{equation}
{1\over \tau} = W_D + {W_R \over (1-W_R)}={W_D\over \langle p \rangle} \label{eq:tau}
%\vspace*{-0.2cm}
\end{equation}
where $W_{D}$ and $W_{R}$ are the thermal activated probabilities defined by 
the Eqs. (\ref{eq:wd})-(\ref{eq:wr}) and calculated for $T_n=T$. In other 
terms, Eq. (\ref{eq:tau}) implies that at a given temperature, both 
$\langle p \rangle$ and $\tau$ only depend on $E_D$ and $E_R$. This equation 
can thus be inverted to determine $E_D$ and $E_R$ as functions of 
$\langle p \rangle$ and $\tau$. These considerations can be generalized to 
the case of a multi-species network by giving the criterion for a convenient 
choice of the activation energies. To this purpose, first we choose a 
reference value of the substrate temperature, $T_{ref}$, and a set of values 
of the average defect fraction, $\langle p_i \rangle$, corresponding to the 
different resistor species (where $p_i \equiv N_{brok,i}/N_{tot,i}$  with 
$i=1,...,N_{spec}$). Second, by considering the correlation times, $\tau_i$, 
associated with the fluctuations \cite{nota} of $p_i$, we choose the values of
the activation energies $E_{D,i}$ and $E_{R,i}$ in such a way to obtain a 
logarithmic distribution of $\tau_i$ inside a sufficiently wide interval: 
$\tau_i \in [\tau_{min},\tau_{max}]$. The different $\tau_i$ are then 
attributed to the different resistor species by adopting the criterion 
that increasing values of $\tau_i$ are paired with increasing values of 
$r_{0,i}$. Alternative choices are of course possible. 

All the results reported here are obtained by applying a bias current 
$I=1.0$ (A) to a network of sizes $75 \times 75$ made by $N_{spec}=15$ 
resistor species and by choosing $r_{min}=0.5$ $\Omega$ and $r_{max}=1.5$ 
$\Omega$. Furthermore we take: $T_{ref}=300$ K, 
$\langle p_i \rangle \approx 0.25 \  \ \forall i$, 
$\tau_{min} \approx 2$ and $\tau_{max} \approx 5 \times 10^5$ 
(where times are expressed in units of iterative steps). We underline 
that the choice made above for the values of $T_{ref}$, $\langle p_i \rangle$,
$\tau_{min}$ and $\tau_{max}$, determines once for all (i.e. for all the 
temperatures) the set of activation energies $E_{D,i}$ and $E_{R,i}$. 
In the present case $E_{D,i}$ and $E_{R,i}$ are in the range 
$58 \div 375$ meV and $37 \div 346$ meV, respectively. In fact, 
the values of $E_{D,i}$ and $E_{R,i}$, together with $r_{0,i}$ and $N_{spec}$,
determine the intrinsic properties of the material under test. Therefore, when 
performing a simulation at temperatures different from $T_{ref}$, the
values of $\tau_i$ recalculated from Eq. (\ref{eq:tau}) are the input 
parameters of the simulation, while the values of $\langle p_i \rangle$ only 
serve as a check of consistency with those obtained directly from the output 
of the simulation. As a general trend, by increasing the temperature the range
of values of $\tau_i$ becomes more narrow and the values of 
$\langle p_i \rangle $ increase progressively.

The time evolution of the network is then obtained by Monte Carlo simulations 
according to the iterative procedure described in details in 
Ref. [\onlinecite{pre_fnl}]. The auto-correlation functions and the power 
spectral densities of the resistance fluctuations are calculated by analyzing 
stationary signals $R(t)$ consisting of $2 \times 10^6$ records.

\section{Results}
The resistance evolution of a multi-species network calculated at a
temperature of $300$ K is reported in Fig. 1. The inset displays a
small part of the same evolution over an enlarged time scale. 
The co-existence of different characteristic time scales in the $R(t)$ signal
is clearly shown in Fig. 1, where the long relaxation time associated 
with the achievement of the steady state, $\tau_{rel}$, co-exists with the 
shorter times characterizing the resistance fluctuations displayed in the 
inset. For comparison, we report in Fig. 2 the time evolution of the 
resistance of a single-species network obtained at $T=300$ K by the 
SBRN model. In this case, the values of the activation energies ($E_D=350$ meV
and $E_R=310$ meV) are chosen to give a relaxation time comparable with that 
of the signal in Fig. 1. The behavior of the resistance is now determined by 
a single time scale ($\tau \approx \tau_{rel}$) and thus it is less noisy
and essentially flat over time scales shorter than $\tau$. This is 
emphasized by the inset of Fig. 2 where the stochastic signal resembles that
of a few level system (it must be noted that the vertical scale of the inset 
in Fig. 2 is significantly enhanced with respect to that of Fig. 1). 

\begin{figure}[bth]
%\vspace*{4.5cm}
%\begin{center}
\includegraphics[height=.25\textheight]{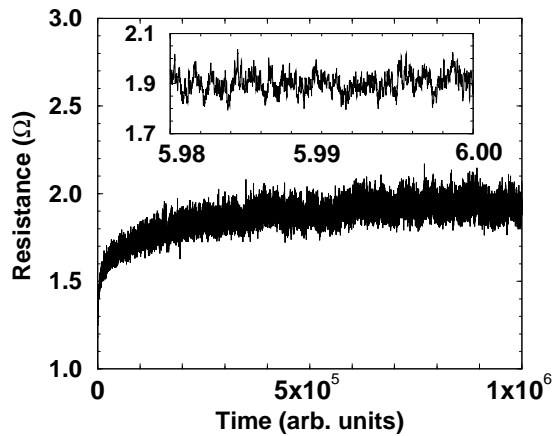}
\caption{\label{fig:epsart}
Resistance evolution of a multi-species network (MSN model)
calculated at 300 K. The resistance is espressed in Ohm and the time  
in iterative steps. The inset highlights the resistance fluctuations 
on an enlarged time scale. In particular, the time units are divided 
by a factor $10^{-5}$.}
%\end{center}
\end{figure}

%\vspace*{4.5cm}
\begin{figure}[bth]
%\begin{center}
\includegraphics[height=.25\textheight]{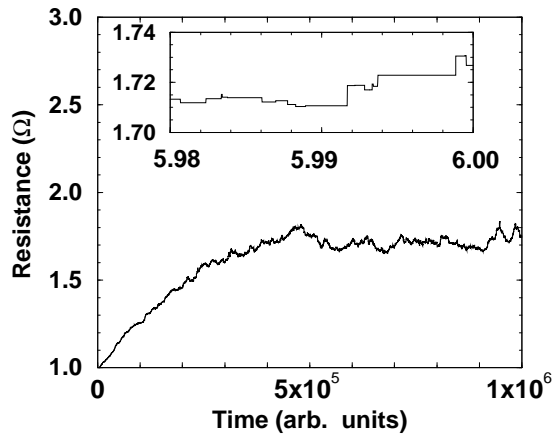}
\caption{\label{fig:epsart} 
Resistance evolution of a single-species network (SBRN model)
calculated at 300 K. The resistance is espressed in Ohm and the time  
in iterative steps. The inset displays the resistance fluctuations on 
the same enlarged time scale of the inset in Fig. 1. (the vertical 
scales of the insets in the two figures are different).}
%\end{center}
\end{figure}
%\vspace*{-0.2cm}

\begin{figure}[bth]
%\vspace*{-1.0cm}
%\begin{center}
\includegraphics[height=.25\textheight]{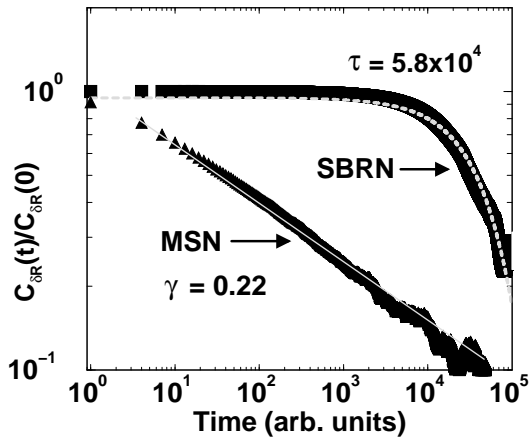}
%\vspace*{-0.2cm}
\caption{\label{fig:epsart}
Auto-correlation functions of the resistance fluctuations 
calculated for a multi-species network (MSN model, black triangles) and for a
single-species network (SBRN model, black squares). Both functions are
obtained at 300 K. The solid and short dashed grey lines show the best-fit
respectively with a power-law of exponent $\gamma=0.22$ and with
an exponential function with correlation time $\tau=5.8 \times 10^4$. 
The time is expressed in iterative steps.}
%\end{center}
\end{figure}
\begin{figure}[bth]
%\vspace*{-1.3cm}
%\begin{center}
\includegraphics[height=.29\textheight]{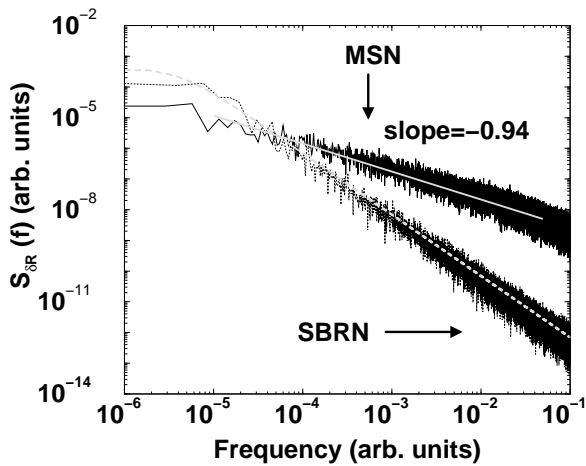}
%\vspace*{-0.4cm}
\caption{\label{fig:epsart}
Power spectral density of the resistance fluctuations at 300 K
calculated for the multi-species network (MSN model, solid line) and for the 
single-species network (SBRN model, dotted line). The grey solid line shows 
the best-fit to the MSN spectrum with a power-law of slope -0.94. The grey 
dashed curve represents the best-fit with a Lorentzian to the SBRN spectrum.}
%\end{center}
\end{figure}

Figure 3 reports the auto-correlation functions of the resistance fluctuations 
on a log-log plot. More precisely, the auto-correlation function corresponding 
to the multi-species network (MSN model) is shown by black triangles while 
that corresponding to the single-species network (SBRN model) is 
reported by black squares. The solid and short dashed lines represent the 
best-fits to the two $C_{\delta R}$ functions carried out respectively 
with a power-law and with an exponential law. The fitting procedure confirms 
the exponential decay of the correlations in the resistance fluctuations of 
the single-species network, described by Eq.~(\ref{eq:cor_exp}), and  
points out the existence of long-term correlations in the resistance 
fluctuations of the multi-species-network, characterized by a power-law 
decay of the auto-correlation function:
\begin{equation}
C_{\delta R}(t) \sim  t^{-\gamma}  \label{eq:power_law}
\end{equation}
with $0 < \gamma <1$. In particular, for the case of Fig. 3 we have found
a value $\gamma=0.22$ for the correlation exponent. We stress that the
above expression of the auto-correlation function implies a divergence of
the correlation time, as can be easily seen by considering the following
general expression of the correlation time \cite{bunde_pa2003}:
\begin{equation}
\tau = \int_{0}^{\infty} {C_{\delta R}(t) \over C_{\delta R}(0)}dt
\end{equation} \label{eq:tau_def}
The power spectral densities of the resistance fluctuations calculated 
at 300 K by the MSN model and by the SBRN model are displayed in Fig. 4. 
The grey solid line represents the best-fit with a power-law to the MSN 
spectrum. The grey dashed curve is the best-fit with a Lorentzian to the SBRN 
spectrum. The corner frequency of the Lorentzian, $f_c=4.0 \times 10^{-6}$ 
(arbitrary units), is consistent with the correlation time reported in 
Fig. 3 and obtained by the best-fit of the corresponding auto-correlation 
function. Thus, Fig. 4 shows that at 300 K the resistance fluctuations of the 
multi-species network exhibit a power spectral density scaling as $1/f^\alpha$
over several decades of frequency, with a value $\alpha=0.94$. This result is 
a consequence of the envelope of the different time scales associated with the
different resistor species. 

Now, we will analyze the effect of the temperature on the resistance
fluctuations of a multi-species network. To this purpose, Fig. 5 displays the 
resistance evolutions calculated at increasing temperatures: $T=400$ K 
(lower curve) and $T=600$ K (upper curve). Already the qualitative comparison 
of the $R(t)$ signals in Figs. 1 and 5, shows that at increasing temperature 
there is a significant growth of both the average resistance and the 
variance of the resistance fluctuations (we will discuss in detail the 
behavior of these quantities in the following). Moreover, these figures
point out a drastic reduction of the relaxation time at increasing temperature 
(more than one order of magnitude when $T$ rises from 300 K to 400 K). 

\begin{figure}[bth]
%\vspace*{-0.6cm}
%\begin{center}
\includegraphics[height=.27\textheight]{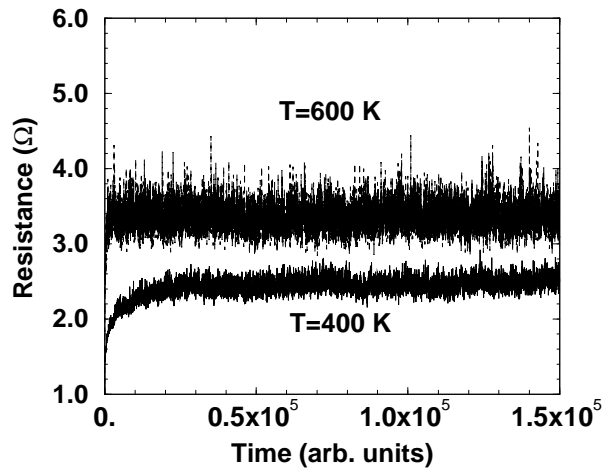}
%\vspace*{-0.1cm}
\caption{\label{fig:epsart}
Resistance evolution of a multi-species network at 400 K and 600 K. 
The resistance is espressed in Ohm and the time in iterative steps.}
%\end{center}
\end{figure}

\begin{figure}[bth]
%\vspace*{4.5cm}
%\begin{center}
\includegraphics[height=.25\textheight]{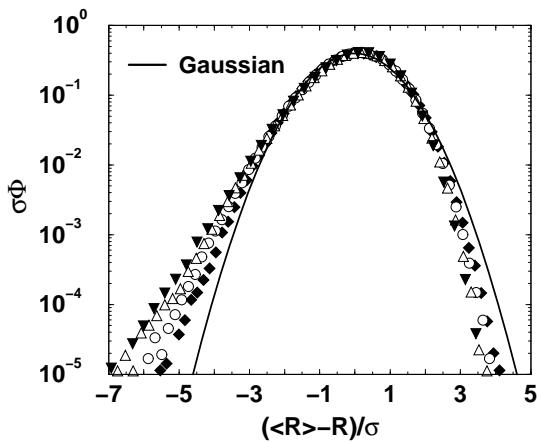}
%\vspace*{-0.4cm}
\caption{\label{fig:epsart}
Normalized probability density functions of the resistance 
fluctuations of a multi-species network calculated at 300 K (full diamonds), 
400 K (open circles), 500 K (open up-triangles) and 600 K (full 
down-triangles); $\sigma$ is the root-mean-square deviation from the average 
resistance value. The solid black curve shows the normalized Gaussian 
distribution.}
%\end{center}
\end{figure}

The temperature is also found to affect the distribution of the resistance 
fluctuations by increasing its skewness, as shown in Fig. 6, which reports on 
a lin-log plot the probability density function (PDF, here denoted by $\phi$),
of the distribution of the resistance fluctuations. A normalized 
representation has been adopted for convenience (here $\sigma$ is the root 
mean square deviation from the average resistance). In Fig. 6, the different 
symbols represent the PDFs calculated in the steady state of the multi-species
network at 300 K (full diamonds), 400 K (open circles), 500 K 
(open up-triangles) and 600 K (full down-triangles). For comparison, the 
normalized Gaussian distribution has been also reported as a solid line. 
The figure shows a significant non-Gaussianity of the resistance fluctuations,
which becomes stronger at high temperatures, meaning that the system is 
approaching failure conditions. The distribution of the resistance 
fluctuations and the role played on the non-Gaussianity by the size, 
shape and disorder of the network has been investigated in Refs. 
[\onlinecite{pen_ng,pen_ng_fn04}], where the link with the universal 
distribution of fluctuations of Bramwell, Holdsworth and Pinton 
\cite{bramwell,clusel} has been discussed. 
 
Figure 7 reports the auto-correlation functions of the resistance fluctuations
of a multi-species network calculated at $400$ K and $600$ K. For comparison, 
the $C_{\delta R}$ function obtained at $300$ K (already shown in Fig. 3) 
has been drawn again in Fig. 7 with the solid grey line still giving the
best-fit with a power-law. The dashed grey curves instead represent the 
best-fit to the $C_{\delta R}$ functions at $400$ K and $600$ K with the 
expression:

\begin{equation}
C_{\delta R} (t)=C_0t^{-h} \exp(-t/u)  \label{eq:power_exp}
\end{equation}

The values of the best-fit parameters are: $C_0=1.13$, $h=0.30$ and 
$u=1.42 \times 10^4$ for $T = 400$ K and $C'_0=0.965$, $h'=0.46$ and 
$u'=5.57 \times 10^2$ for $T = 600$ K. As pointed out by Fig. 7,
the fit to $C_{\delta R}$ with Eq.~(\ref{eq:power_exp}) is very satisfying. 
We thus conclude that at high temperatures the auto-correlation function of 
the resistance fluctuations of the multi-species network is well described 
by a power-law with an exponential cut-off. Such kinds of mixed decays of 
the correlations, non-exponential and non power-law, are often found in the 
transition of a complex system from short-term correlated to long-term 
correlated regimes \cite{sornette,pen_epj}. It should be noted that 
for $u \rightarrow \infty$ Eq.~(\ref{eq:power_exp}) becomes a 
power-law while for $h \rightarrow 0$ it describes an exponential decay. 

Actually, Fig. 7 makes evident a strong reduction of the correlation time
of the fluctuations when the temperature increases. This result can be
easily understood in terms of Eq.~(\ref{eq:tau}), by considering the
thermally activated expressions of the breaking and recovery probabilities:
in fact the increase of temperature implies a significant reduction of the 
ratio $\tau_{min}/\tau_{max}$. Therefore the interval $[\tau_{min},\tau_{max}]$
where the $\tau_i$ are distributed becomes more and more narrow, destroying 
the power-law decay of correlations and giving rise to an exponential tail. 
It must be noted that, for similar reasons, when the temperature decreases 
below the reference value, $T<T_{ref}$, the ratio $\tau_{min}/\tau_{max}$ 
strongly increases and the correlations in the resistance fluctuations keep 
their power-law decay over wider time scales (for example, when $T=200$ K 
the ratio $\tau_{min}/\tau_{max}$ becomes  $7.5 \times 10^7$). $T_{ref}$ 
is thus closely related to the transition temperature, $T^*$, from a 
long-term correlated behavior i.e. a behavior characterized by a divergent 
correlation time and occurring for $T<T^*$, to a behavior characterized by a 
finite correlation time and occurring for $T>T^*$.   

The correlation time of the resistance fluctuations can be directly estimated 
at $T > T_{ref}$ by considering Eq. (7) and making use of 
Eq.~(\ref{eq:power_exp}). We obtain the following expression of $\tau$ 
in terms of the best-fit parameters of the auto-correlation function:
\begin{equation}
\tau = u^{1-h} \Gamma(1-h) \label{eq:tau_fit}
\end{equation}
where $\Gamma$ is the Gamma function. The values of $\tau$ calculated in 
such a way are reported in Fig. 8 as a function of the temperature. The
figure shows a strong increase of $\tau$ at decreasing temperatures, when 
$T$ approches $T_{ref}$, in fact for this value of temperature $\tau$ 
is expected to diverge consistently with the power-law behavior of the 
auto-correlation function. To highlight the dependence of $\tau$ on the 
temperature, we report in Fig. 9 a log-log plot of the correlation time as a 
function of the difference $T-T^*$, where the value of the transition 
temperature $T^*=306$ K is determined by a best-fit to the data in Fig. 8 
with the power-law: $\tau \sim (T-T^*)^{-\theta}$. The fitting curve
is displayed in Fig. 9 by the dashed straight line. We find a value 
$\theta=2.7$ for the exponent. Therefore Fig. 9 shows that the dependence of 
the correlation time on temperature is well described by a power of  $T-T^*$, 
where the transition temperature $T^* \approx T_{ref}$.

\begin{figure}[bth]
%\vspace*{-0.5cm}
%\begin{center}
\includegraphics[height=.25\textheight]{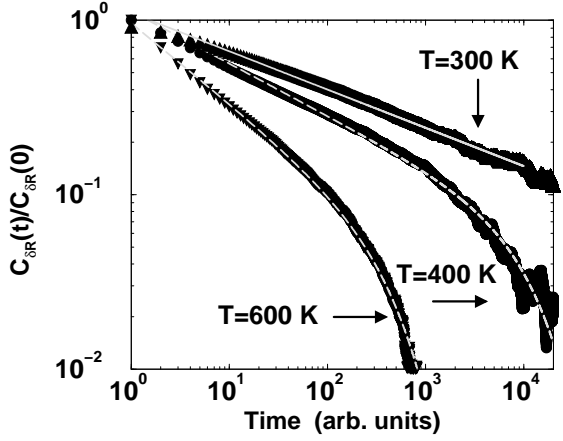}
%\vspace*{-0.5cm}
\caption{\label{fig:epsart}
Auto-correlation functions of the resistance fluctuations of 
a multi-species network calculated at different temperatures. The solid grey
curve shows the best-fit with a power-law to the auto-correlation function 
at 300 K (the same of Fig. 3). The dashed grey curves display the best-fit 
to the auto-correlation functions at 400 and 600 K with the function:
$C(t)=C_0t^{-h} \exp[-t/u]$ (see the text for the values of the fit 
parameters). The time is expressed in iterative steps.} 
%\end{center}
\end{figure}

\begin{figure}[bth]
%\vspace*{-0.7cm}
%\begin{center}
\includegraphics[height=.25\textheight]{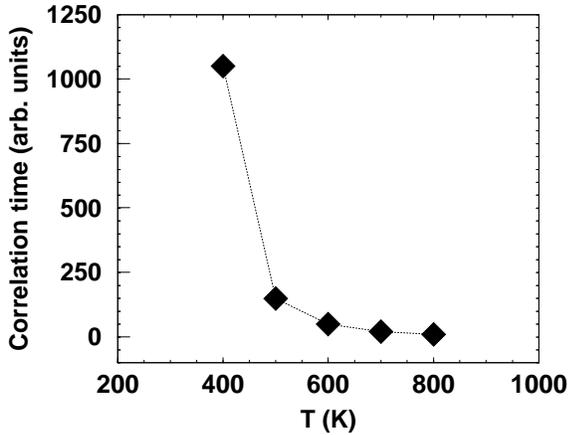}
%\vspace*{-0.4cm}
\caption{\label{fig:epsart}
Correlation time of the resistance fluctuations as a function of 
the temperature. The time is expressed in iterative steps and the 
temperature in K. The dotted line connecting the diamonds is a guide 
to the eyes.}
%\end{center}
\end{figure}

\begin{figure}[bth]
%\vspace*{-0.7cm}
%\begin{center}
\includegraphics[height=.25\textheight]{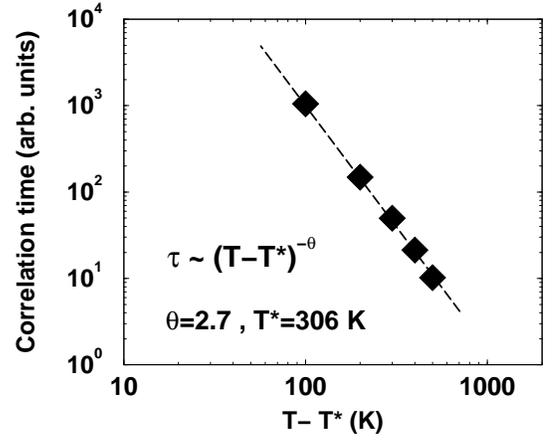}
%\vspace*{-0.2cm}
\caption{\label{fig:epsart}
Correlation time of the resistance fluctuations as a function of 
the difference $T-T^*$. The time is expressed in iterative steps and the 
temperature in K. The value of $T^*$ is reported in the figure. 
The dashed line shows the fit with a power-law of exponent $\theta=2.7$.}
%\end{center}
\end{figure}

\begin{figure}[bth]
%\vspace*{4.7cm}
%\begin{center}
\includegraphics[height=.27\textheight]{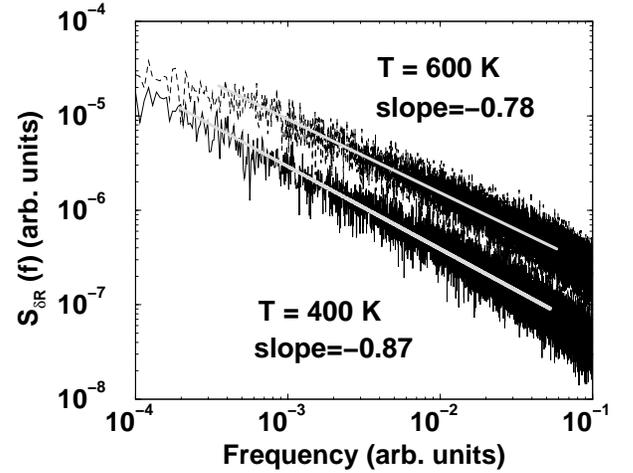}
%\vspace*{-0.5cm}
\caption{\label{fig:epsart}
Power spectral density of the resistance fluctuations of the 
multi-species network at $T=400$ K and $T=600$ K. The grey lines show the 
best-fit with power-laws of slopes -0.87 and -0.78, respectively. }
%\end{center}
\end{figure}

%\vspace*{-1.5cm} 
The power spectral densities of the resistance fluctuations calculated at the
temperatures of $400$ K and  $600$ K are reported in Fig. 10. Here the grey 
lines represent the best-fit with power-laws of slopes -0.87 and -0.78, 
respectively. Thus we find that the power spectrum keeps the $1/f^{\alpha}$ 
form also at high temperatures. However, the value of the exponent ${\alpha}$ 
decreases to values well below unity when $T>T^*$. This dependence of 
${\alpha}$ on temperature has been studied since a long time 
\cite{review,alers}. A qualitative description of this dependence is provided 
by the well known Dutta-Horn relation \cite{review}. On the other hand, many 
experiments have pointed out a strong dependence of the detailed behavior 
of $\alpha(T)$ on the particular material \cite{review,upon05,alers}. 
In this respect, it should be noted that a decrease of the $\alpha$ exponent 
from $\alpha \approx 1$ and $\alpha \approx 0.8-0.5$ is frequently observed 
in the experiments \cite{review,upon05,alers} at intermediate temperatures 
above 100-200 K, similarly to our case where this decrease occurs above 
$T_{ref} \approx T^* \approx 300$ K. It should be noted, however, that 
in our model $T_{ref}$ merely plays the role of an input parameter which can
be adapted to fit the experiments. Furthermore, we stress that the monotonic 
decrease of ${\alpha}$ for  $T>T^*$ is obtained here in the linear regime 
of currents, i.e. neglecting Joule heating effects. Actually, these effects, 
whose importance also depends on temperature can give rise to a more 
complicated behavior of ${\alpha}$ versus temperature.

Now we consider the dependence on temperature of the average resistance and 
of the variance of the resistance fluctuations. We call $R_0$ the average 
value of the resistance in the limit $T=0$. This value represents the 
resistance of a network free of broken resistors but in any case disordered 
(due to the presence of different resistor species). Thus $R_0$ not only 
depends on the network size and on the set of values $\{r_{0,i}\}$ with 
$i=1,...N_{spec}$, but also on the particular random distribution of the 
$r_{0,i}$ within the network. Figure 11 displays the difference 
$\langle R \rangle - R_0$ as a function of the temperature. A log-log 
representation is adopted here for convenience. The dashed straight line 
represents the best-fit of the data with the expression: 
$\langle R \rangle = R_0 +A T^{\gamma}$, with $\gamma=0.83$, 
$A=1.35 \times 10^{-2} \ \Omega/K^{\gamma}$ and $R_0= 0.52\ \Omega$. We can 
see that this expression describes rather well the behavior of the average 
resistance. 

\begin{figure}[bth]
%\vspace*{-0.5cm}
%\begin{center}
\includegraphics[height=.25\textheight]{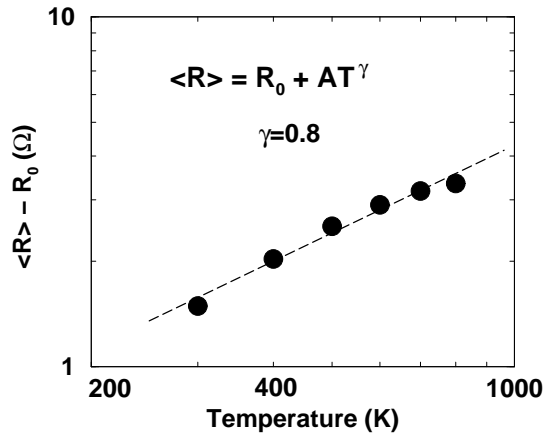}
%\vspace*{-0.4cm}
\caption{\label{fig:epsart}
Variation of the average resistance as a function of the temperature. 
The resistance is espressed in Ohm and the temperature in K. The dashed 
straight line represents the best-fit with the expression reported in the 
figure.}
%\end{center}
\end{figure}

\begin{figure}[bth]
%\vspace*{-0.5cm}
%\begin{center}
\includegraphics[height=.25\textheight]{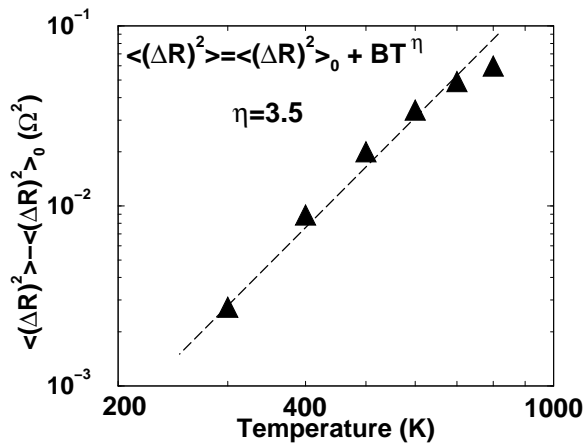}
\caption{\label{fig:epsart}
Change of the variance of the resistance fluctuations as a function 
of the temperature. The variance is espressed in $\Omega^2$ and the  
temperature in K. The dashed straight line shows a best-fit with the 
expression reported in the figure.}
%\end{center}
\end{figure}

\begin{figure}[bth]
%\vspace*{-1.0cm}
%\begin{center}
\includegraphics[height=.25\textheight]{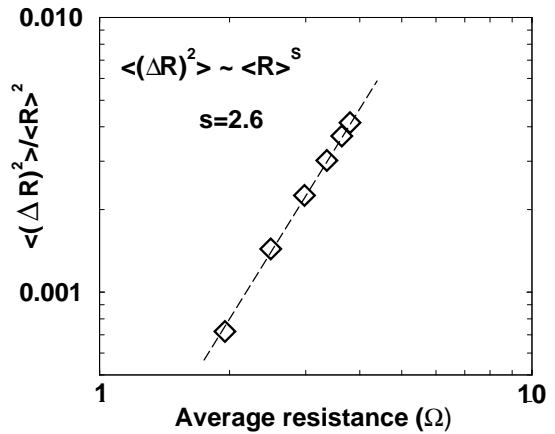}
%\vspace*{-0.4cm}
\caption{\label{fig:epsart}
Relative variance of the resistance fluctuations as a function of the
average resistance (this last is expressed in Ohm). The dashed line shows a 
best-fit with a power-law of exponent $s=2.6$.}
%\end{center}
\end{figure}

The change with the temperature of the variance of the resistance fluctuations
is reported in Fig. 12. Here $\langle (\Delta R)^2 \rangle_0$ is the limit 
value of the variance at $T=0$. This value mainly depends on the network size 
\cite{prl_stat} and it vanishes for $N \rightarrow \infty$. Again, a log-log 
representation is used in this figure. The dashed straight line displays 
the best-fit with the expression: 
$\langle (\Delta R)^2 \rangle = \langle (\Delta R)^2 \rangle _0 + B T^{\eta}$
which is found to provide a good description of the data for $\eta=3.5$,
$B= 7.32 \times 10^{-12} \ \Omega^2/K^{\eta}$ and 
$\langle (\Delta R)^2 \rangle_0 = 5.0 \times 10^{-7} \ \Omega^2$ .

Finally, Fig. 13 shows in a log-log plot the relative variance of the 
resistance fluctuations as a function of the average resistance. In this case 
the dashed line represents the best-fit with a power-law of exponent $2.6$. 
In other terms, Fig. 13 shows that: 
\begin{equation}
\langle (\Delta R)^2 \rangle /\langle R \rangle^2 \ \sim 
\langle R \rangle^s \label{eq:relvar}
\end{equation}
where $s=2.6$. To be consistent with the behaviors of the average resistance 
and of the variance of the resistance fluctuations reported in Figs. 11-12, 
the following scaling relation should hold within the exponents 
$s$, $\eta$ and $\gamma$: $s=\eta/\gamma-2$. By using the previously 
reported values of $\eta$ and $\gamma$ we obtain for $s$ the value $2.4$,
in agreement (within the error) with the result $s=2.6$ obtained by directly
considering the dependence of 
$\langle (\Delta R)^2 \rangle /\langle R \rangle^2$  on $\langle R \rangle$. 
We underline that both, the behavior of the relative variance and the value 
of $s$, are perfectly consistent with the results reported in 
Ref. [\onlinecite{prl_stat}]. This agreement confirms and points out the fact 
that the model discussed here directly generalizes the results of 
Ref. [\onlinecite{prl_stat}] to resistors characterized by $1/f^{\alpha}$ 
resistance noise. 

\vspace*{-0.5cm}
\section{Conclusions}
\vspace*{-0.5cm}
We have developed a stochastic model to investigate the $1/f^{\alpha}$ 
resistance noise in disordered materials. More precisely, we have considered 
the resistance fluctuations of a thin resistor with granular structure in a 
non-equilibrium stationary state. This system has been modeled as a 
two-dimensional network made by different species of elementary resistors. The
steady state of this multi-species network is determined by the competition 
among different thermally activated and stochastic processes of breaking and 
recovery of the elementary resistors. The network properties have then been 
studied by Monte Carlo simulations as a function of the temperature and in 
the linear regime of the external bias. By a suitable choice of the values of 
the parameters, the model gives rise to resistance fluctuations with different
power spectra, thus providing a unified approach to the study of materials 
exhibiting either Lorentzian noise or $1/f^{\alpha}$ noise. In particular, 
by increasing the temperature in the range from $300$ to $800$ K the $\alpha$ 
exponent decreases from values near to 1 down to values around $0.5$, 
in qualitative agreement with experimental findings available from the
literature \cite{review,upon05,alers}. This behavior, indicative of a drastic 
reduction of the correlation time of the resistance fluctuations, has been 
also analysed in terms of the auto-correlation functions. These last 
progressively change at increasing temperature, going from a slow, power-law 
decay of the correlations at $300$ K to a faster but non-exponential decay 
at higher temperatures.

\section*{ACKNOWLEDGMENTS}
%\vspace*{-1.0cm}
Support from MIUR cofin-05 project 
"Strumentazione elettronica integrata per lo studio di variazioni 
conformazionali di proteine tramite misure elettriche" is acknowledged. 

%\vspace*{-0.5cm}

%\end{multicols}


\begin{thebibliography}{10}
\bibitem{nota} We take the correlation time associated with the fluctuations 
of the defect fraction, $\tau_p$ comparable with that corresponding to the
fluctuations of the resistance, $\tau$, i.e. we take  $\tau_p \approx \tau$.
Our simulations confirm this assumption.
\end{thebibliography}
\end{document}